\documentclass[%
 reprint,
 amsmath,amssymb,
 showpacs,
 prd
]{revtex4-1}

\usepackage{amsmath}
\usepackage{graphicx}
\usepackage{dcolumn}
\usepackage{bm}
\usepackage{verbatim}
\usepackage{array}
\usepackage{amsmath}
\usepackage{lineno}
\usepackage{threeparttable}
\usepackage[bookmarksnumbered, pdfstartview=FitH,colorlinks,urlcolor=blue, citecolor=blue,linkcolor=blue,]{hyperref}
\usepackage{setspace}
\usepackage{textcomp}
\usepackage{multirow}

\usepackage{mathtools}
\begin{document}
\preprint{APS/123-QED}

\title{\boldmath Sensitivity study of $\bar{K}_1(1270)$ decay dynamics using four $D\to \bar{K}_1(1270)(\to \bar K\pi\pi)e^+\nu$ decay channels}

\author{Ying'ao Tang$^{1}$}
\email{yingaotang@whu.edu.cn}
\author{Liang Sun$^{1}$}
\email{sunl@whu.edu.cn}
\author{Panting Ge$^{2}$}
\author{Menghao Wang$^{1}$}
\affiliation{%
        $^{1}$School of Physics and Technology, Wuhan University\\
        $^{2}$School of Physics, Henan Normal University\\
}%

\date{\today}


\begin{abstract}
A sensitivity study for the measurement of $\bar{K}_1(1270)$ decay modes is performed using semileptonic $D$-meson decays. The BESIII experiment is taken as a case study, where a simultaneous analysis of $\bar{K}_1(1270)$ decays to the four thee-body final states $K^-\pi^+\pi^-$, $K^-\pi^+\pi^0$, $K_S^0\pi^+\pi^-$, and $K_S^0\pi^-\pi^0$ is presented and a model-independent determination of \mbox{$\mathcal{B}(\bar{K}_1(1270)\to \bar K\pi\pi)$}, without requiring detailed knowledge of intermediate resonant contributions, is proposed. 
\end{abstract}


\maketitle

\section{Introduction}
\noindent 

The strange axial-vector mesons offer interesting possibilities for the study of quantum chromodynamics in the non-perturbative regime. Due to the presence of a strange quark with mass greater than the up and down quark masses, SU(3) symmetry is broken so that the $^3P_1$ and $^1P_1$ states mix with each other to construct the mass eigenstates, $\bar{K}_1(1200)$ and $\bar{K}_1(1400)$, by the mixing angle $\theta_{\bar{K}_1}$~\cite{Cheng:2017pcq}. The mixing angle $\theta_{\bar{K}_1}$ plays a crucial role in determining the theoretical calculations, such as the helicity form factors and branching fractions (BFs) for semileptpnic $D$-meson decays into strange axial-vector mesons~\cite{Momeni:2020zrb,Momeni:2019uag}.

Semileptonic charm decays, induced by the quark-level process $c\to se^+\nu_e$, are  predominantly mediated by pseudoscalar ($K$) and vector ($K^*(892)$) mesons, i.e., contain a kaon and at most one pion in the final-state hadronic systems~\cite{Isgur:1988gb,Scora:1995ty}. However, semileptonic charm decays to higher-multiplicity final states are expected to proceed predominantly via the axial–kaon system~\cite{Hatanaka:2008gu} and are therefore strongly suppressed. The $D\to \bar{K}\pi\pi e^+\nu_e$ decays provide a unique opportunity to study the properties and interactions of $\bar{K}_1(1270)$ and $\bar{K}_1(1400)$ mesons in a clean environment, without any additional hadrons in the final states. Such studies can lead to a better determination of $\theta_{\bar{K}_1}$, as well as more precise measurements of the masses and widths of the $\bar{K}_1$ mesons, all of which currently carry large uncertainties~\cite{ParticleDataGroup:2024cfk}. Furthermore, by exploiting the measured properties of $D\to \bar{K}_1(1270) \ell^+\nu_\ell$ and $B\to \bar{K}_1(1270)\gamma$ decays, the photon polarization in $b\to s\gamma$ can be determined  without considerable theoretical ambiguity~\cite{Wang:2019wee,Bian:2021gwf}. 
Charge-conjugate decays are implied throughout the paper. 

The BFs of $\bar K_1(1270)$ decays to different two-body final states of $\bar K\rho,\ \bar K^{*}_0(1430)\pi,\ \bar K^*(892)\pi,\ \bar K \omega, \bar Kf_0(1370)$ reported by the Particle Data Group (PDG)~\cite{ParticleDataGroup:2024cfk}
are mostly based on a study of the $K^-\pi^+\pi^-$ system conducted in a $K^- p\to K^-\pi^-\pi^+ p$ scattering experiment in 1981~\cite{CNTR}, combined with a recent BESIII measurement of the branching ratio ${\cal B}(\bar K(1270)\to\bar K^*(892)\pi)/{\cal B}(\bar K(1270)\to\bar K\rho)$ in the $D_s^+\to K^-K^+\pi^+\pi^0$ decay~\cite{BESIII:2021qfo}. All these BFs possesses large uncertainties, that lead to $\sim$20\% uncertainties on the $\bar K_1(1270)\to \bar K \pi\pi$ BFs~\cite{BESIII:2024awg}, becoming a bottleneck for precise BF measurements on any decays with $\bar K_1(1270)$ as intermediate particles. 

Although not used by the PDG for the BF averages, there are still a number of other measurements on the $\bar K_1(1270)$ decays. Based on an amplitude analysis of the decay $B^+\to J/\psi K^+\pi^+\pi^-$, the Belle collaboration found the BFs of  $\bar{K}_1(1270)\to \bar{K}\rho, \bar K\omega$, and  $\bar{K}^*(892)\pi$  to be generally consistent with the PDG averages within two standard deviations, while the measured BF of $\bar{K}_1(1270)\to K_0^*(1430)\pi$ is significantly smaller~\cite{Belle:2010wrf}. Later measurements of the BF ratio   $\alpha\equiv\frac{\mathcal{B}(\bar{K}_1(1270)\to \bar K^*\pi)}{\mathcal{B}(\bar{K}_1(1270)\to \bar{K}\rho)} $, where \mbox{$\mathcal{B}(\bar{K}_1(1270)\to \bar K^*\pi)=\mathcal{B}(\bar{K}_1(1270)\to \bar{K}^*(1430)\pi)$} $+\mathcal{B}(\bar{K}_1(1270)\to \bar{K}^*(892)\pi)$, yield different results depending on the decay channels used~\cite{CLEO:2012beo,BESIII:2017jyh,LHCb:2017swu,dArgent:2017gzv}, whereas they are expected to be identical under the narrow width approximation for the $\bar{K}_1(1270)$ meson assuming $CP$ conservation in strong decays~\cite{Guo:2018orw}. 

The BESIII collaboration, through performing separate studies of the four hadronic systems $K^-\pi^+\pi^-$~\cite{BESIII:2019eao}, $K^-\pi^+\pi^0$~\cite{BESIII:2021uqr}, $K_S^0\pi^+\pi^-$ and $K_S^0\pi^-\pi^0$~\cite{BESIII:2024ieo}, reported the first observations of semielectronic $D$-meson decays involving a $\bar{K}_1(1270)$ and measured their BFs based on the assumed $\bar{K}_1(1270)$ decays. In addition, quite recently, an amplitude analysis of the \mbox{$D^0\to K^-\pi^+\pi^-e^+\nu_e$} and $D^+\to K^-\pi^0\pi^-e^+\nu_e$ decays has been performed~\cite{BESIII:2025hdt} with the larger $\psi(3770)$ dataset corresponding to an integrated luminosity of $20.3~\text{fb}^{-1}$. The measured BFs are summarized in Tab.~\ref{branching_fractions}. 
In light of these measurements, in this work, a model-independent method is proposed to determine the BFs of $\bar{K}_1(1270)$ decays through a simultaneous analysis of signal yields from the four decay modes $D^0\to K^-\pi^+\pi^-e^+\nu_e$, $D^+\to K^-\pi^+\pi^0e^+\nu_e$, $D^0\to K^0_S\pi^-\pi^0e^+\nu_e$, and $D^+\to K^0_S\pi^+\pi^-e^+\nu_e$. With this method, the feasibility of measuring BFs of $\mathcal{B}(\bar{K}_1(1270)\to \bar K^*\pi)$, $\mathcal{B}(\bar{K}_1(1270)\to \bar{K}\rho)$ and $\mathcal{B}(\bar K_1(1270)\to \bar K\pi\pi)$, based on the current 20.3~fb$^{-1}$ $\psi(3770)$ data sample from BESIII~\cite{BESIII:2024lbn}, is explored. The projected precisions on the BFs are also evaluated using pseudo-experiments. 

\begin{table*}[t]
    \centering
    \caption{Summary of measured BFs and corresponding integrated luminosities ($\int {\cal L}\text{d}t$) for four $D$-meson semileptonic decay modes in Ref.~\cite{BESIII:2019eao, BESIII:2021uqr,BESIII:2024ieo,BESIII:2025hdt}. The first and second uncertainties are statistical and systematic, respectively.  For BFs of \mbox{$D\to \bar{K}_1(1270)e^+\nu_e$} decays, the third uncertainties originate from the assumed BFs of $\bar{K}_1(1270)$ decays~\cite{ParticleDataGroup:2024cfk}.}
    {\renewcommand{\arraystretch}{1.3}
    \begin{tabular}{lcccc}
    \hline
    Decay mode & Signal yield &$\mathcal{B}(D\to \bar K\pi\pi e^+\nu)\times 10^{-4}$ & $\mathcal{B}(D\to \bar{K}_1(1270)e^+\nu)\times 10^{-3}$ &$\int {\cal L}\text{d}t$\\ \hline

    \multirow{2}{*}{$D^0\to K^-\pi^+\pi^-e^+\nu_e$} & $109\pm13$& $(3.95\pm 0.41^{+0.31}_{-0.52}$ ) & $(1.09\pm 0.13^{+0.09}_{-0.16} \pm0.12$ ) & $2.93~\text{fb}^{-1}$ \\
     & $731\pm 35$& $(3.20\pm 0.20\pm0.20)$ & $(1.02\pm 0.06\pm 0.06 \pm0.03$ ) & $20.3~\text{fb}^{-1}$ \\ 
    \cline{2-5}
    
    \multirow{2}{*}{$D^+\to K^-\pi^+\pi^0e^+\nu_e$} & $120\pm 13$& $(10.6\pm 1.2 \pm 0.8$ )  &  $\left(2.30\pm 0.26^{+0.18}_{-0.21} \pm 0.25\right)$  & $2.93~\text{fb}^{-1}$\\ 
     & $1270\pm 56$& $(12.70\pm 0.60 \pm 0.40$ )  &  $\left(2.27\pm 0.11 \pm0.07 \pm 0.07\right)$ & $20.3~\text{fb}^{-1}$\\ 
    \cline{2-5}
    
    $D^0\to K^0_S\pi^-\pi^0e^+\nu_e$ & $17\pm5$& $(1.69^{+0.53}_{-0.46 }\pm0.15$ ) & $\left(1.05_{-0.28}^{+0.33} \pm 0.12 \pm 0.12\right)$  & $2.93~\text{fb}^{-1}$\\
    \cline{2-5}
    
    $D^+\to K^0_S\pi^+\pi^-e^+\nu_e$ &  $20\pm6$&$(1.47^{+0.45}_{-0.40 }\pm0.14$ )  &  $\left(1.29_{-0.35}^{+0.40} \pm 0.18 \pm 0.15\right)$  & $2.93~\text{fb}^{-1}$\\
    \hline
    \end{tabular}}
    \label{branching_fractions}
\end{table*}






\section{Formalism}
Table~\ref{branching_fractions} lists the experimentally measured BFs \mbox{$\mathcal{B}(D\to \bar{K}_1(1270)e^+\nu_e)$}, which depend on the assumed decay BFs of  $\bar{K}_1(1270)$. In the measurements~\cite{BESIII:2019eao, BESIII:2021uqr,BESIII:2024ieo, BESIII:2025hdt}, the BF $\mathcal{B}(D \to \bar K\pi\pi e^+\nu_e)$ can be expressed as the product of \mbox{$\mathcal{B}(D \to \bar{K}_1(1270) e^+\nu_e)$} and \mbox{$\mathcal{B}(\bar{K}_1(1270) \to \bar K\pi\pi)$}~\cite{formu}, where $\mathcal{B}(\bar{K}_1(1270)\to \bar K\pi\pi)$ represents the sum BFs of $\bar{K}_1(1270)$ decay into $K\pi\pi$ final states:
\begin{equation}
    \mathcal{B}(\bar{K}_1(1270)\to \bar K\pi\pi)= \sum_{i}f_i\mathcal{B}^i(\bar{K}_1(1270)\to f), \label{eq2}
\end{equation}
where $f_i$ is the square of Clebsch–Gordan coefficients corresponding to the $i^{\text{th}}$ decay mode of $\bar{K}_1(1270)$: \mbox{$\bar{K}^*(1430)\pi$}, $~\bar{K}^*(892)\pi ,~\bar{K}\rho,~\bar K\omega$. The last decay mode is ignored hereafter due to smallness of the product ${\cal B}(\bar{K}_1(1270)\to \bar K\omega)\times {\cal B}(\omega\to\pi^+\pi^-)$. 
Concerning the completeness of the $\bar K_1(1270)$ decays, the sum $\mathcal{B}_{\text{3body}}$ is defined as

\begin{equation}
\begin{split}
    \mathcal{B}_{\text{3body}} =& \mathcal{B}(\bar{K}_1(1270) \to \bar{K}\rho)
    + \mathcal{B}(\bar{K}_1(1270) \to \bar K^*\pi) \\
    & + \mathcal{B}(\bar{K}_1(1270) \to \bar{K}f_0)\\=&1-\mathcal{B}(\bar{K}_1(1270) \to \bar K\omega)=(0.89\pm0.02)\%, 
\end{split}
\end{equation}
which is determined by $\mathcal{B}(\bar{K}_1(1270) \to \bar K\omega)$~\cite{ParticleDataGroup:2024cfk}. 

A transition variable, which are directly related to the BFs of different signal reconstruction modes, and can be determined experimentally in a straight-forward fashion, is defined as:
\begin{equation}
    \beta^{-1}\equiv1-\frac{\mathcal{B}(K^-_1(1270)\to K^-\pi^+\pi^-)}{\mathcal{B}(\bar{K}_1^0(1270)\to K^-\pi^+\pi^0)}. \label{betadefine}
\end{equation}

By inserting Eq.~(\ref{betadefine}) into the expression for $\alpha$ and defining the ratio $\delta_\alpha\equiv\frac{\mathcal{B}(\bar{K}_1(1270)\to \bar{K}f_0)}{\mathcal{B}(\bar{K}_1(1270)\to \bar{K}\rho)}$, the BF ratio $\alpha$ is expressed as:
\begin{equation}
\alpha=\frac{3}{4}[\beta(1-3\delta_\alpha)-2] \label{alphadefine}.
\end{equation}
The BFs of $\bar{K}_1(1270)$ decay can then be expressed as:
\begin{equation}
\begin{aligned}&\mathcal{B}(K_1^-(1270)\rightarrow K^{-} \pi^{+} \pi^{-})\\&=
\mathcal{B}_{\text{3body}} \cdot \frac{3+4 \alpha+9\delta_\alpha}{9(1+\alpha+\delta_\alpha)}\\
&=\frac{4}{3}\mathcal{B}_{\text{3body}}\cdot\frac{\beta-3\delta_\alpha\beta+3\delta_\alpha-1}{(3\beta-9\delta_\alpha\beta+4\delta_\alpha-2)},\end{aligned}\label{BFs1}
\end{equation}

\begin{equation}
    \begin{aligned}&\mathcal{B}(\bar{K}_1^0(1270)\rightarrow K^{-} \pi^{+} \pi^0)\\ &=\mathcal{B}_{\text{3body}} \cdot \frac{6+4 \alpha}{9(1+\alpha+\delta_\alpha)} \\&=\frac{4}{3}\mathcal{B}_{\text{3body}}\cdot\frac{\beta-3\delta_\alpha\beta}{(3\beta-9\delta_\alpha\beta+4\delta_\alpha-2)}.\end{aligned}\label{BFs2}
\end{equation}
Since $\delta_\alpha$ is related to $\mathcal{B}(\bar{K}_1(1270)\to \bar{K}\rho)$,  it is necessary to eliminate its dependence to make the measurement model-independent. Introducing the shorthand $r_{f_0}=\frac{\mathcal{B}(\bar{K}_1(1270)\to \bar{K}f_0)}{\mathcal{B}_{\text{3body}}-\mathcal{B}(\bar{K}_1(1270)\to \bar{K}f_0)}=(3.5\pm2.3)\%$ \cite{ParticleDataGroup:2024cfk} into Eq.~(\ref{alphadefine}) can better gauge the related uncertainty:

\begin{equation}
    \begin{aligned}
        \alpha&=\frac{3}{4}[\beta(1-r_{f_0}(1+\alpha)-2]\\ &=\frac{3(\beta-3r_{f_0}\beta-2)}{9r_{f_0}\beta+4}=\frac{3\beta-2}{9r_{f_0}\beta+4}-1,
    \end{aligned}
\end{equation}
which leads to an updated expression for $\delta_\alpha$ :

\begin{equation}
    \begin{aligned}
        \delta_{\alpha}=r_{f_0}(1+\alpha)=\frac{r_{f_0}(3\beta-2)}{9r_{f_0}\beta+4}.
    \end{aligned}
\end{equation}

By eliminating $\alpha$ and substituting $\delta_\alpha$, the BFs  \mbox{$\mathcal{B}(K^-_1(1270)\rightarrow K^{-} \pi^{+} \pi^{-})$ and $\mathcal{B}(\bar{K}^0_1(1270)\rightarrow K^{-} \pi^{+} \pi^0)$} in Eq.~(\ref{BFs1}-\ref{BFs2}) can be expressed as:

\begin{equation}
\label{eq9}
\begin{aligned}   &\mathcal{B}(K_1^-(1270)\rightarrow K^{-} \pi^{+} \pi^{-})
\\ & =\frac{4}{3} \mathcal{B}_{\text {3body }} \cdot \frac{(\beta-1)(1-3 \delta_\alpha)}{3 \beta(1-3 \delta_\alpha)+4 \delta_\alpha-2}
\\ &= \frac{4}{3} \mathcal{B}_{\text {3body }} \cdot \frac{(\beta-1)\cdot (4+6 r_{f_0})}{4(3\beta-2)(r_{f_0}+1)},
\end{aligned}
\end{equation}

\begin{equation}
\label{eq10}
\begin{aligned}&\mathcal{B}(\bar{K}_1^0(1270)\rightarrow K^{-} \pi^{+} \pi^0)
\\& =\frac{4}{3} \mathcal{B}_{\rm 3body } \cdot \frac{\beta(1-3 \delta_\alpha)}{3 \beta(1-3\delta_\alpha)+4 \delta_\alpha-2}
\\ &= \frac{4}{3} \mathcal{B}_{\text {3body }} \cdot \frac{\beta\cdot (4+6 r_{f_0})}{4(3\beta-2)(r_{f_0}+1)}. 
\end{aligned}
\end{equation}
 In Eqs.~\ref{eq9} and \ref{eq10}, the parameter \(\beta\) serves as the sole free variable in the formulation, while $\mathcal{B}_{\text {3body }}$ 
 and $r_{f_0}$ rely on external inputs of ${\cal B}(\bar K_1(1270)\to \bar K \omega)$ and ${\cal B}(\bar K_1(1270)\to \bar K f_0(1370))$ from the PDG. The value \(\beta\) can be directly determined from experimental data through a fit to extract the corresponding signal yields. Once \(\beta\) is extracted, all other physical observables—including the BFs and related quantities, can be derived from it, as they are explicit functions of \(\beta\). This framework thus provides a consistent and model-independent approach, where all derived parameters are fully constrained by the experimentally determined value of \(\beta\).

\section{Experimental potentials}\label{Sec::2}

\begin{table*}[t]
\centering
\caption{The measured values from the simultaneous fit to one pseudo-dataset statistically matched to the 20.3~$\rm fb^{-1}$ $\psi(3770)$ dataset from BESIII. For each result (``Output'') of this work, the first uncertainty is statistical, the second one is systematic, and the last uncertainty is from the external input of $\mathcal{B}(\bar K_1(1270) \to \bar K\omega)$~\cite{ParticleDataGroup:2024cfk}. The BESIII results~\cite{BESIII:2025hdt} are listed for comparison. }

\begin{threeparttable}
\begin{tabular}{cccc} \hline
Parameters & Input & Output & BESIII results \\
\hline 
$\alpha$ [\%] &  $20.3$ & $22.7\pm 15.0\pm 1.0\pm 0.6$ & $20.3\pm 2.1\pm 8.7$  \\ 

$\mathcal{B}(\bar{K}_1(1270)\rightarrow \bar{K}^*(892)\pi)$ [\%] & $ 15.0$ & $16.5\pm 9.0\pm 0.7\pm 3.5$ & $19.5\pm 1.9\pm 5.2$~\tnote{$\star$} \\
&& &  $10.9\pm1.2\pm3.0$~\tnote{$\dagger$} \\ 
$\mathcal{B}(\bar{K}_1(1270)\rightarrow \bar{K}\rho)$ [\%] &   $ 74.0$ & $72.5\pm 9.0\pm 0.7\pm 3.5$ & $71.8\pm 2.3\pm 23.9$~\tnote{$\star$} \\
&&&$79.3\pm 2.0\pm25.7$~\tnote{$\dagger$} \\ 
$\mathcal{B}(K_1^-(1270)\rightarrow K^{-} \pi^{+} \pi^{-})$ [\%] & $31.3$ & $31.5\pm 1.1\pm 0.7\pm 0.4$ & $31.3\pm 0.9$ \\
$\mathcal{B}(\bar{K}_1^0(1270)\rightarrow K^{-} \pi^{-} \pi^0)$ [\%] & $56.0$  & $55.7\pm 2.1\pm 1.3\pm 0.8$  &$56.0\pm 2.7$\\
$\mathcal{B}(D^0\to K_1^-(1270)e^+\nu_e)$ [$\times 10^3$]& $1.02$&$1.01\pm0.05\pm 0.02\pm 0.01$ & $1.02\pm0.06\pm 0.06\pm 0.03$ \\
$\mathcal{B}(D^+\to \bar{K}_1^0(1270)e^+\nu_e)$ [$\times 10^3$]& $2.27$& $2.29\pm0.10\pm 0.05\pm 0.01$ & $2.27\pm0.11\pm 0.07\pm 0.07$ \\ 
\hline
\end{tabular}
\begin{tablenotes}\footnotesize
\item[$\star$] From the channel of $D^0\to K^-\pi^+\pi^-e^+\nu_e$;
\item[$\dagger$] From the channel of $D^+\to K^-\pi^+\pi^0 e^+\nu_e$.
\end{tablenotes}
\end{threeparttable}
    \label{fitresult}
\end{table*}

The BESIII collaboration has individually measured $D^0\to K^-\pi^+\pi^-e^+\nu_e$, $D^+\to K^-\pi^0\pi^-e^+\nu_e$, $D^0\to K^0_S\pi^0\pi^-e^+\nu_e$ and $D^+\to K^0_S\pi^+\pi^-e^+\nu_e$ decays~\cite{BESIII:2019eao, BESIII:2021uqr,BESIII:2024ieo}, with the double-tag method \cite{MARK-III:1985hbd,MARK-III:1987jsm} and the $\psi(3770)$ dataset corresponding to an integrated luminosity of $2.93~\text{fb}^{-1}$. A combined analysis of the $D^0\to K^-\pi^+\pi^-e^+\nu_e$ and $D^+\to K^-\pi^0\pi^-e^+\nu_e$ decays has been performed~\cite{BESIII:2025hdt} with the larger $\psi(3770)$ dataset corresponding to an integrated luminosity of $20.3~\text{fb}^{-1}$. In this section, a sensitivity study is performed by simultaneously fitting across the four decay modes, to determine the BFs of $\bar{K}_1(1270)$ decays, and the BF ratio $\alpha$, at the same time.

One-dimensional pseudo-datasets of $M_{\rm miss}^2$ are generated for the decay modes of $D^0\to K^-\pi^+\pi^-e^+\nu_e$ and $D^{0,+}\to K_S^0\pi^-\pi^{0,+}e^+\nu_e$. Here $M_{\rm miss}^2$ is the missing mass square $M_{\rm miss}^2 \equiv E^2_{\rm miss}/c^4-|\vec p_{\rm miss}|^2/c^2$, with $E_{\rm miss}$ and $\vec p_{\rm miss}$ being the total energy and momentum of all missing particles in the event, respectively. For the decay mode of $D^+\to K^-\pi^+\pi^0e^+\nu_e$, as the distribution of $U_{\rm miss}\equiv E_{\rm miss}-|\vec p_{\rm miss}|c$ was used instead for signal yield extraction in Ref~\cite{BESIII:2019eao}, the signal and background shapes of $M_{\rm miss}^2$ from the mode of $D^{0}\to K_S^0\pi^-\pi^{0}e^+\nu_e$ are used as approximations. 
The expected signal yields of the decays $D^0\to K^-\pi^+\pi^-e^+\nu_e$, $D^+\to K^-\pi^0\pi^-e^+\nu_e$, $D^0\to K^0_S\pi^0\pi^-e^+\nu_e$ and $D^+\to K^0_S\pi^+\pi^-e^+\nu_e$, based on the $20.3~\text{fb}^{-1}$ $\psi(3770)$ data, are estimated with
\begin{equation}
\begin{aligned}
&N(D\to \bar{K}\pi\pi e^+\nu_e) \\ &= 2N(D\bar D) \times \sum_i\mathcal{B}_i^{\rm ST}\varepsilon^{i}_{\rm DT}\times \mathcal{B}(D
\to \bar{K}\pi\pi e^+\nu_e)
\end{aligned}
\end{equation}
where $N(D\bar D)$ denotes the total number of produced $D\bar{D}$ pairs~\cite{BESIII:2024lbn}, $\mathcal{B}_i^{\rm ST}$ is the BF of the $i^{\rm th}$ tag mode, and $\varepsilon_{i}^{\rm DT}$ is the double-tag efficiency. The summation runs over the same tag modes as those used in Refs.~\cite{BESIII:2019eao,BESIII:2021uqr, BESIII:2024ieo}, with the values of $\varepsilon_{i}^{\rm DT}$ also assumed to be the same as in Refs.~\cite{BESIII:2019eao,BESIII:2021uqr, BESIII:2024ieo}.  

The background events are generated from the background probability density functions that were previously determined from Monte Carlo (MC) simulations in \cite{BESIII:2021uqr, BESIII:2024ieo}. The estimated yields of combinatorial and $D\to \bar K\pi\pi\pi$ peaking backgrounds are all scaled by a factor of seven to account for the smaller datasets used in Refs.~\cite{BESIII:2019eao,BESIII:2021uqr, BESIII:2024ieo}.

To extract the parameters of interest, a simultaneous unbinned maximum likelihood fit is performed across the four pseudo-datasets. The probability density functions modeling the signal and background components are adopted from Refs.~\cite{BESIII:2021uqr, BESIII:2024ieo}, that are determined from MC simulations. During the fit, the signal and combinatorial background yields are allowed to float, while the yields of the peaking backgrounds are fixed to their generated values. 

\begin{figure}
    \centering
    \includegraphics[width=1\linewidth]{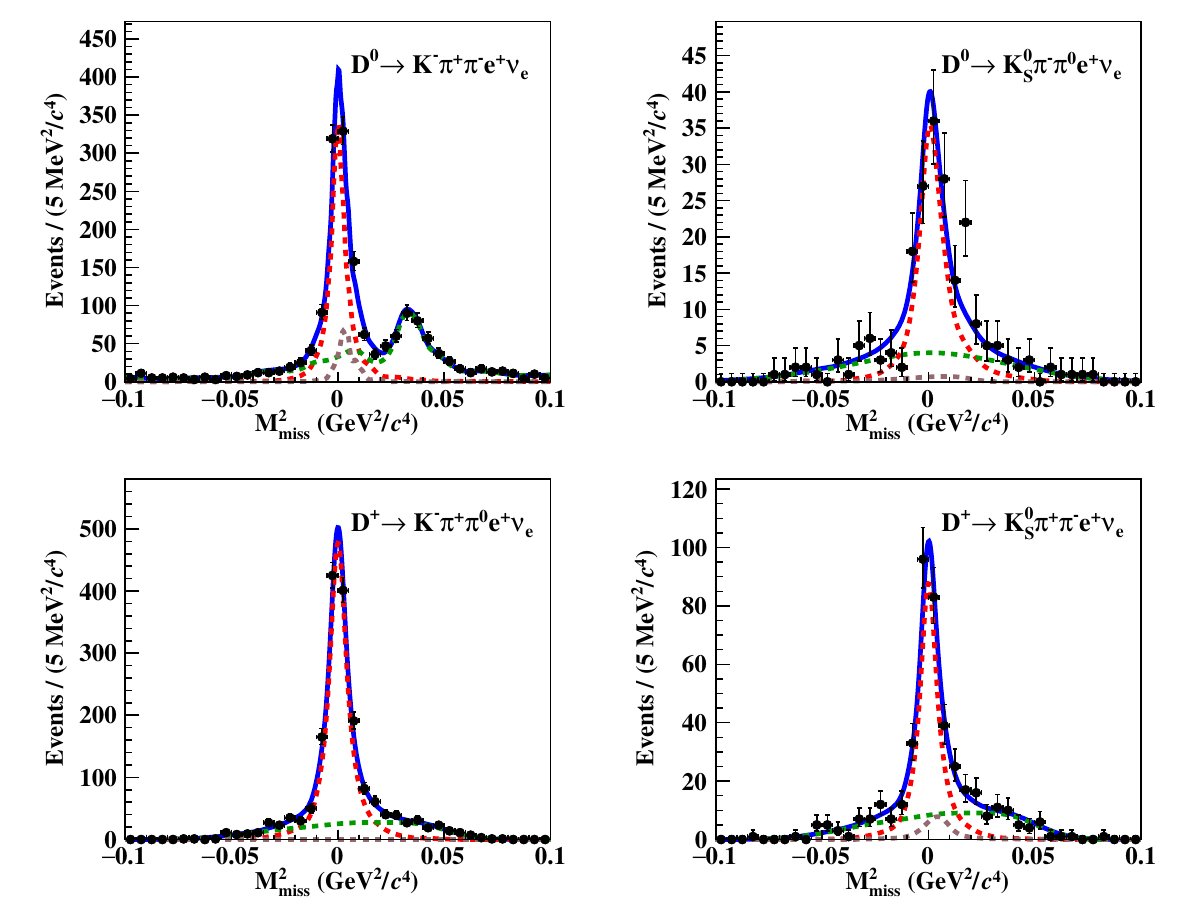}
    \caption{Simultaneous fit to the $M_{\rm miss}^2$  distributions of the pseudo-datasets. The points with error bars are the pseudo-data, the red dashed lines represent the signal shapes, and the green dashed lines represent the combinatorial background shapes. The brown dashed lines denote the peaking backgrounds from $D\to \bar K\pi\pi\pi$ decays. }
    \label{fitprojection}
\end{figure}

To minimize systematic effects from common sources such as luminosity, tagging and tracking efficiencies, the parameter $\beta$ is reformulated in terms of ratios of signal yields:
\begin{equation}\label{ratio1}
\begin{aligned}
\beta_{D^0}^{-1}&=1-\frac{\mathcal{N}(K^-\pi^+\pi^-e^+\nu_e)}{\mathcal{N}(K^-\pi^+\pi^0e^+\nu_e)}\\ &=1-\frac{\mathcal{N}(K^-\pi^+\pi^-e^+\nu_e)}{\frac{\mathcal{N}(K_S^0\pi^-\pi^0e^+\nu_e)}{\mathcal{B}(K_S^0\to \pi^+\pi^-)/2}},
\end{aligned}
\end{equation}

\begin{equation}\label{ratio2}
        \begin{aligned}\beta_{D^+}^{-1}&=1-\frac{\mathcal{N}(K^-\pi^+\pi^-e^+\nu_e)}{\mathcal{N}(K^-\pi^+\pi^0e^+\nu_e)}\\ &=1-\frac{\frac{\mathcal{N}(K^0_S\pi^+\pi^-e^+\nu_e)}{\mathcal{B}(K_S^0\to \pi^+\pi^-)/2}}{\mathcal{N}(K^-\pi^+\pi^0e^+\nu_e)}.\end{aligned}
\end{equation}
where $\mathcal{N}$ denotes the efficiency-corrected signal yield for each decay mode. By assuming $\beta_{D^0}=\beta_{D^+}$, the averaged value of $\beta$ can be extracted from the simultaneous fit to the pseudo-datasets across the four decay modes, from which the other observables can subsequently be determined using Eq.~\ref{eq9} and \ref{eq10}. The one-dimensional fit projections to the $M^2_{\rm miss}$ distributions for the four decays are shown in Fig.~\ref{fitprojection} and the fit results are summarized in Tab.~\ref{fitresult}.  

\begin{figure}[h]
    \centering
    \includegraphics[width=0.7\linewidth]{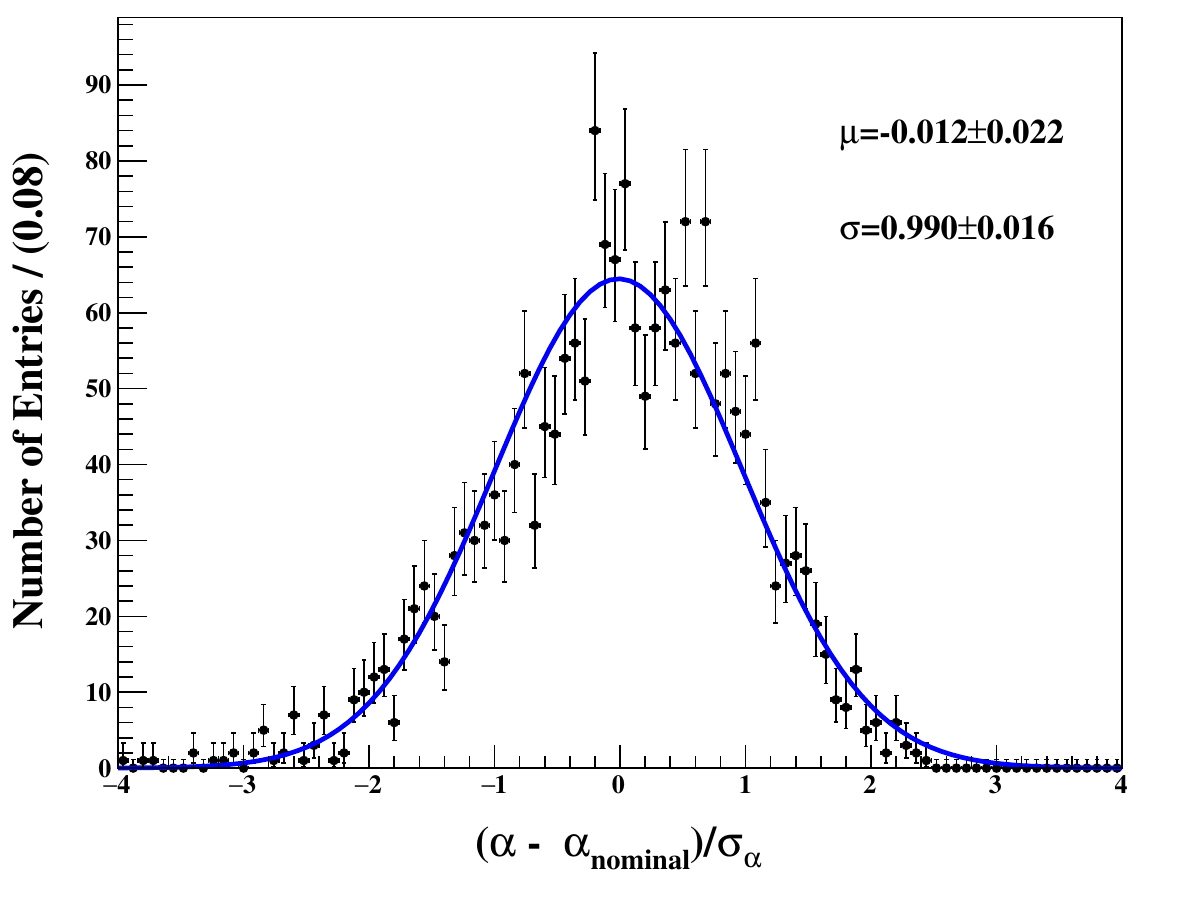}
    \caption{The $\alpha$ pull distribution, illustrating the difference between the reconstructed value $\alpha'$ and the input value $\alpha_{\rm nominal}$ normalized by the estimated uncertainty $\sigma_{\alpha}$.}
    \label{fig:pull}
\end{figure}

A total of 2000 pseudo-experiments are performed to assess potential biases introduced by the fit model. The resulting distribution of the pulls, defined as $\frac{\alpha_{\rm fit}-\alpha_{\rm nominal}}{\sigma_{\rm fit}}$, where $\alpha_{\rm fit}$ ($\sigma_{\rm fit}$) is the fitted $\alpha$ central value (uncertainty) in each pseudo-experiment, is shown in Fig.~\ref{fig:pull} and is consistent with a normal distribution, indicating that the fit model is unbiased in determining $\alpha$.

Concerning potential sources of systematic uncertainties in the measurement, the double-tag method ensures that most of the uncertainties arising from the tag side cancel. The uncertainties associated with the tracking and particle-identification efficiencies of $e^+$ and charged pions mostly cancel in the ratios in Eqs.~\ref{ratio1} and \ref{ratio2}. The uncertainty of the $\pi^0$ and $K_S^0$ reconstruction efficiencies is 1\% \cite{BESIII:2024tpv,BESIII:2024awg}. The systematic uncertainty associated with this is evaluated by applying a Gaussian constraint to the efficiency parameters during the fit, yielding a relative uncertainty of 3.8\%. Similarly, the uncertainty originating from the assumed input branching fractions ($\mathcal{B}_{\rm 3~body}$ and $r_{f_0}$) is estimated by applying Gaussian constraints to these input parameters, which contributes an additional 2.7\%. Adding these independent sources in quadrature results in a total conservative estimate of~5\%


Compared to the BESIII's amplitude analysis based on the 20.3 fb$^{-1}$ $\psi(3770)$ dataset~\cite{BESIII:2025hdt}, the expected statistical uncertainties on $\alpha$ and the BFs on $\bar{K}_1(1270)\to \bar{K}\rho,\bar K^*\pi$ in this work are larger. Because this work does not exploit the full kinematic information of the $D\to \bar K\pi\pi e^+ \nu_e$ decay (e.g., angular and $q^2$ distributions). Still, by taking into account the systematic uncertainties, this method is able to achieves significantly improved precision for ${\cal B}(\bar{K}_1(1270)\to \bar{K}\rho,\bar K^*\pi)$. With this method, the expected precisions on BFs on $\bar{K}_1(1270)\to K^-\pi^{+,0}\pi^-$ are comparable to the BESIII results, while the input uncertainties on ${\cal B}(D\to \bar{K}_1(1270) e^+\nu_e)$ are considerably reduced. 

\section{Summary}

In this work, a sensitivity study is performed to evaluate the feasibility of measuring the absolute BF $\mathcal{B}(\bar{K}_1(1270)\to \bar K\pi\pi)$ and the ratio $\alpha=\mathcal{B}(\bar{K}_1(1270)\to  \bar{K}^*\pi)/\mathcal{B}(\bar{K}_1(1270)\to \bar{K}\rho)$. A model-independent approach to study $\bar{K}_1(1270)$ decay is proposed via simultaneously extracting signal yields from the four $K\pi\pi$ final states via fitting. The study demonstrates that a systematic uncertainty of around 5\% can be obtained with the current  $\psi(3770)$ data sample (20.3~fb$^{-1}$) from the BESIII experiment, providing a significant improvement over previous results~\cite{BESIII:2025hdt}. 

Not relying on specific signal decay models, this combined analysis yields substantially lower systematic uncertainties for $\mathcal{B}(\bar{K}_1(1270)\to \bar K^*\pi)$, $\mathcal{B}(\bar{K}_1(1270)\to \bar{K}\rho)$, and their ratio $\alpha$, while providing a robust, model-independent validation of existing amplitude analysis results. Furthermore, these results are able to lay the groundwork for high-precision probes of axial-vector meson structure and decay dynamics  and will become increasingly advantageous with larger datasets from the Super Tau-Charm Factory, where the statistical uncertainties are expected be reduced by at least one order of magnitude~\cite{Achasov:2023gey,Fan:2021mwp}.

\bibliography{refs.bib}

@article{Isgur:1988gb,
    author = "Isgur, Nathan and Scora, Daryl and Grinstein, Benjamin and Wise, Mark B.",
    title = "{Semileptonic B and D Decays in the Quark Model}",
    reportNumber = "UTPT-88-12",
    doi = "10.1103/PhysRevD.39.799",
    journal = "Phys. Rev. D",
    volume = "39",
    pages = "799--818",
    year = "1989"
}

@article{Scora:1995ty,
    author = "Scora, Daryl and Isgur, Nathan",
    title = "{Semileptonic meson decays in the quark model: An update}",
    eprint = "hep-ph/9503486",
    archivePrefix = "arXiv",
    reportNumber = "CEBAF-TH-94-14",
    doi = "10.1103/PhysRevD.52.2783",
    journal = "Phys. Rev. D",
    volume = "52",
    pages = "2783--2812",
    year = "1995"
}

@article{Hatanaka:2008gu,
    author = "Hatanaka, Hisaki and Yang, Kwei-Chou",
    title = "{$K_1(1270)-K_1(1400)$ Mixing Angle and New-Physics Effects in $B \to K_1 l^+ l^-$ Decays}",
    eprint = "0808.3731",
    archivePrefix = "arXiv",
    primaryClass = "hep-ph",
    doi = "10.1103/PhysRevD.78.074007",
    journal = "Phys. Rev. D",
    volume = "78",
    pages = "074007",
    year = "2008"
}

@article{Cheng:2017pcq,
    author = "Cheng, Hai-Yang and Kang, Xian-Wei",
    title = "{Branching fractions of semileptonic $D$ and $D_s$ decays from the covariant light-front quark model}",
    eprint = "1707.02851",
    archivePrefix = "arXiv",
    primaryClass = "hep-ph",
    doi = "10.1140/epjc/s10052-017-5170-5",
    journal = "Eur. Phys. J. C",
    volume = "77",
    number = "9",
    pages = "587",
    year = "2017",
    note = "[Erratum: Eur.Phys.J.C 77, 863 (2017)]"
}

@article{Momeni:2020zrb,
    author = "Momeni, S.",
    title = "{Helicity form factors for $D_{(s)} \rightarrow A \ell \nu $ process in the light-cone QCD sum rules approach}",
    eprint = "2004.02522",
    archivePrefix = "arXiv",
    primaryClass = "hep-ph",
    doi = "10.1140/epjc/s10052-020-8084-6",
    journal = "Eur. Phys. J. C",
    volume = "80",
    number = "6",
    pages = "553",
    year = "2020"
}

@article{Momeni:2019uag,
    author = "Momeni, S. and Khosravi, R.",
    title = "{Semileptonic $D_{(s)} \to A \ell^+ \nu$ and nonleptonic $D\to K_1(1270,1400) \pi$ decays in LCSR}",
    eprint = "1903.00860",
    archivePrefix = "arXiv",
    primaryClass = "hep-ph",
    doi = "10.1088/1361-6471/ab35d0",
    journal = "J. Phys. G",
    volume = "46",
    number = "10",
    pages = "105006",
    year = "2019"
}

@article{ParticleDataGroup:2024cfk,
    author = "Navas, S. and others",
    collaboration = "Particle Data Group",
    title = "{Review of particle physics}",
    doi = "10.1103/PhysRevD.110.030001",
    journal = "Phys. Rev. D",
    volume = "110",
    number = "3",
    pages = "030001",
    year = "2024"
}

@article{Wang:2019wee,
    author = "Wang, Wei and Yu, Fu-Sheng and Zhao, Zhen-Xing",
    title = "{Novel Method to Reliably Determine the Photon Helicity in  $B\to K_{1}\gamma$}",
    eprint = "1909.13083",
    archivePrefix = "arXiv",
    primaryClass = "hep-ph",
    doi = "10.1103/PhysRevLett.125.051802",
    journal = "Phys. Rev. Lett.",
    volume = "125",
    number = "5",
    pages = "051802",
    year = "2020"
}

@article{Bian:2021gwf,
    author = "Bian, Lingzhu and Sun, Liang and Wang, Wei",
    title = "{Up-down asymmetries and angular distributions in $D\to K_1(K\pi\pi)l^+\nu_{l}$}",
    eprint = "2105.06207",
    archivePrefix = "arXiv",
    primaryClass = "hep-ph",
    doi = "10.1103/PhysRevD.104.053003",
    journal = "Phys. Rev. D",
    volume = "104",
    number = "5",
    pages = "053003",
    year = "2021"
}

@article{BESIII:2019eao,
    author = "Ablikim, Medina and others",
    collaboration = "BESIII",
    title = "{Observation of the Semileptonic $D^+$ Decay into the $\bar K_1(1270)^0$ Axial-Vector Meson}",
    eprint = "1907.11370",
    archivePrefix = "arXiv",
    primaryClass = "hep-ex",
    doi = "10.1103/PhysRevLett.123.231801",
    journal = "Phys. Rev. Lett.",
    volume = "123",
    number = "23",
    pages = "231801",
    year = "2019"
}

@article{BESIII:2021uqr,
    author = "Ablikim, Medina and others",
    collaboration = "BESIII",
    title = "{Observation of $D^0\to K_1(1270)^- e^+\nu_e$}",
    eprint = "2102.10850",
    archivePrefix = "arXiv",
    primaryClass = "hep-ex",
    doi = "10.1103/PhysRevLett.127.131801",
    journal = "Phys. Rev. Lett.",
    volume = "127",
    number = "13",
    pages = "131801",
    year = "2021"
}

@article{BESIII:2024ieo,
    author = "Ablikim, Medina and others",
    collaboration = "BESIII",
    title = "{Observation of the semileptonic decays $ {D}^0\to {K}_{\textrm{S}}^0{\pi}^{-}{\pi}^0{e}^{+}{\nu}_{e} $ and $ {D}^{+}\to {K}_{\textrm{S}}^0{\pi}^{+}{\pi}^{-}{e}^{+}{\nu}_{e} $}",
    eprint = "2403.19091",
    archivePrefix = "arXiv",
    primaryClass = "hep-ex",
    doi = "10.1007/JHEP09(2024)089",
    journal = "JHEP",
    volume = "09",
    pages = "089",
    year = "2024"
}

@article{MARK-III:1985hbd,
    author = "Baltrusaitis, R. M. and others",
    collaboration = "MARK-III",
    title = "{Direct Measurements of Charmed d Meson Hadronic Branching Fractions}",
    reportNumber = "SLAC-PUB-3861",
    doi = "10.1103/PhysRevLett.56.2140",
    journal = "Phys. Rev. Lett.",
    volume = "56",
    pages = "2140",
    year = "1986"
}

@article{MARK-III:1987jsm,
    author = "Adler, J. and others",
    editor = "Tran Thanh Van, J.",
    collaboration = "MARK-III",
    title = "{A Reanalysis of Charmed d Meson Branching Fractions}",
    reportNumber = "SLAC-PUB-4291",
    doi = "10.1103/PhysRevLett.60.89",
    pages = "153--160",
    year = "1987"
}

@article{BESIII:2025hdt,
    author = "Ablikim, M. and others",
    collaboration = "BESIII",
    title = "{First Measurement of the Decay Dynamics in the Semileptonic Transition of $D^{+(0)}$ into the Axial-Vector Meson $K_1(1270)$}",
    doi = "10.1103/xj42-xgzf",
    journal = "Phys. Rev. Lett.",
    volume = "135",
    number = "9",
    pages = "091801",
    year = "2025"
}

@article{Belle:2010wrf,
    author = "Guler, H. and others",
    collaboration = "Belle",
    title = "{Study of the $K^+ \pi^+ \pi^-$ Final State in $B^+ \to J/\psi K^+ \pi^+ \pi^-$ and $B^+ \to \psi^\prime K^+ \pi^+ \pi^-$}",
    eprint = "1009.5256",
    archivePrefix = "arXiv",
    primaryClass = "hep-ex",
    reportNumber = "BELLE-PREPRINT-2010-20, KEK-PREPRINT-2010-32",
    doi = "10.1103/PhysRevD.83.032005",
    journal = "Phys. Rev. D",
    volume = "83",
    pages = "032005",
    year = "2011"
}

@article{BESIII:2021qfo,
    author = "Ablikim, Medina and others",
    collaboration = "BESIII",
    title = "{Amplitude analysis and branching fraction measurement of $D_s^+ \to K^-K^+\pi^+\pi^0$}",
    eprint = "2103.02482",
    archivePrefix = "arXiv",
    primaryClass = "hep-ex",
    doi = "10.1103/PhysRevD.104.032011",
    journal = "Phys. Rev. D",
    volume = "104",
    number = "3",
    pages = "032011",
    year = "2021"
}

@book{formu,
title = "{$\mathcal{B}(K^-_1(1270)\to K^-\pi^+\pi^-) = \frac{1}{3}\mathcal{B}_{K_1\to K\rho} + \frac{4}{9}\mathcal{B}_{K_1\to K^*\pi} + \mathcal{B}_{K_1\to K\omega}\times\mathcal{B}_{\omega\to \pi^+\pi^-} + \mathcal{B}_{K_1\to Kf_0}, \mathcal{B}(\bar{K}^0_1(1270)\to K^-\pi^+\pi^0) = \frac{2}{3}\mathcal{B}_{K_1\to K\rho} + \frac{4}{9}\mathcal{B}_{K_1\to K^*\pi}$}"

}

@article{Guo:2018orw,
    author = "Guo, Peng-Fei and Wang, Di and Yu, Fu-Sheng",
    title = "{Strange Axial-vector Mesons in $D$ Meson Decays}",
    eprint = "1801.09582",
    archivePrefix = "arXiv",
    primaryClass = "hep-ph",
    doi = "10.11804/NuclPhysRev.36.02.125",
    journal = "Nucl. Phys. Rev.",
    volume = "36",
    number = "2",
    pages = "125--134",
    year = "2019"
}

@article{CLEO:2012beo,
    author = "Artuso, M. and others",
    collaboration = "CLEO",
    title = "{Amplitude analysis of $D^0\to K^+K^-\pi^+\pi^-$}",
    eprint = "1201.5716",
    archivePrefix = "arXiv",
    primaryClass = "hep-ex",
    reportNumber = "CLNS-11-2082, CLEO-11-08",
    doi = "10.1103/PhysRevD.85.122002",
    journal = "Phys. Rev. D",
    volume = "85",
    pages = "122002",
    year = "2012"
}

@article{BESIII:2017jyh,
    author = "Ablikim, Medina and others",
    collaboration = "BESIII",
    title = "{Amplitude analysis of $D^{0} \rightarrow K^{-} \pi^{+} \pi^{+} \pi^{-}$}",
    eprint = "1701.08591",
    archivePrefix = "arXiv",
    primaryClass = "hep-ex",
    doi = "10.1103/PhysRevD.95.072010",
    journal = "Phys. Rev. D",
    volume = "95",
    number = "7",
    pages = "072010",
    year = "2017"
}

@article{LHCb:2017swu,
    author = "Aaij, Roel and others",
    collaboration = "LHCb",
    title = "{Studies of the resonance structure in $D^{0} \rightarrow K^\mp \pi ^\pm \pi ^\pm \pi ^\mp $ decays}",
    eprint = "1712.08609",
    archivePrefix = "arXiv",
    primaryClass = "hep-ex",
    reportNumber = "LHCB-PAPER-2017-040, CERN-EP-2017-314",
    doi = "10.1140/epjc/s10052-018-5758-4",
    journal = "Eur. Phys. J. C",
    volume = "78",
    number = "6",
    pages = "443",
    year = "2018"
}

@article{dArgent:2017gzv,
    author = "d'Argent, Philippe and Skidmore, Nicola and Benton, Jack and Dalseno, Jeremy and Gersabeck, Evelina and Harnew, Sam and Naik, Paras and Prouve, Claire and Rademacker, Jonas",
    title = "{Amplitude Analyses of $D^0 \to {\pi^+\pi^-\pi^+\pi^-}$ and $D^0 \to {K^+K^-\pi^+\pi^-}$ Decays}",
    eprint = "1703.08505",
    archivePrefix = "arXiv",
    primaryClass = "hep-ex",
    doi = "10.1007/JHEP05(2017)143",
    journal = "JHEP",
    volume = "05",
    pages = "143",
    year = "2017"
}

@article{BESIII:2024lbn,
    author = "Ablikim, Medina and others",
    collaboration = "(BESIII),, BESIII",
    title = "{Measurement of integrated luminosity of data collected at 3.773 GeV by BESIII from 2021 to 2024*}",
    eprint = "2406.05827",
    archivePrefix = "arXiv",
    primaryClass = "hep-ex",
    doi = "10.1088/1674-1137/ad70a0",
    journal = "Chin. Phys. C",
    volume = "48",
    number = "12",
    pages = "123001",
    year = "2024"
}

@article{BESIII:2024tpv,
    author = "Ablikim, Medina and others",
    collaboration = "BESIII",
    title = "{Observation of $D\to a_0(980)\pi$ in the decays $D^0\to\pi^+\pi^-\eta$ and $D^+\to\pi^+\pi^0\eta$}",
    eprint = "2404.09219",
    archivePrefix = "arXiv",
    primaryClass = "hep-ex",
    doi = "10.1103/PhysRevD.110.L111102",
    journal = "Phys. Rev. D",
    volume = "110",
    number = "11",
    pages = "L111102",
    year = "2024"
}

@article{BESIII:2024awg,
    author = "Ablikim, Medina and others",
    collaboration = "BESIII",
    title = "{Analysis of the dynamics of the decay $ {D}^{+}\to {K}_S^0{\pi}^0{e}^{+}{\nu}_e $}",
    eprint = "2408.04422",
    archivePrefix = "arXiv",
    primaryClass = "hep-ex",
    doi = "10.1007/JHEP10(2024)199",
    journal = "JHEP",
    volume = "10",
    pages = "199",
    year = "2024"
}

@article{Achasov:2023gey,
    author = "Achasov, M. and others",
    title = "{STCF conceptual design report (Volume 1): Physics {\&} detector}",
    eprint = "2303.15790",
    archivePrefix = "arXiv",
    primaryClass = "hep-ex",
    doi = "10.1007/s11467-023-1333-z",
    journal = "Front. Phys. (Beijing)",
    volume = "19",
    number = "1",
    pages = "14701",
    year = "2024"
}

@article{Fan:2021mwp,
    author = "Fan, Yu-Lan and Shi, Xiao-Dong and Zhou, Xiao-Rong and Sun, Liang",
    title = "{Feasibility study of measuring $b\rightarrow s\gamma $ photon polarisation in $D^0\rightarrow K_1(1270)^- e^+\nu _e$ at STCF}",
    eprint = "2107.06118",
    archivePrefix = "arXiv",
    primaryClass = "hep-ex",
    doi = "10.1140/epjc/s10052-021-09841-y",
    journal = "Eur. Phys. J. C",
    volume = "81",
    number = "12",
    pages = "1068",
    year = "2021"
}

@article{CNTR,
    author = "Daum, C. and others",
    collaboration = "ACCMOR",
    title = "{Diffractive Production of Strange Mesons at 63-{GeV}}",
    reportNumber = "CERN-EP/81-04",
    doi = "10.1016/0550-3213(81)90114-0",
    journal = "Nucl. Phys. B",
    volume = "187",
    pages = "1--41",
    year = "1981"
}
\end{document}